\documentclass[preprint,letterpaper,prc,showpacs,11pt]{revtex4}
\usepackage{graphicx}
\newcommand{\Tr}{{\rm Tr}}
\newcommand{\sss}{\scriptscriptstyle}
\begin{document}
\preprint{KVI-1632}
\title{Bulk Viscosity in Neutron Stars from Hyperons}
\email{vandalen@kvi.nl; eric.van-dalen@uni-tuebingen.de; dieperink@kvi.nl}
\author{E.N.E. van Dalen \footnote{Current Address: Institut
f$\ddot{\textrm{u}}$r Theoretische Physik, Universit$\ddot{\textrm{a}}$t
T$\ddot{\textrm{u}}$bingen,
Auf der Morgenstelle 14, D-72076 T$\ddot{\textrm{u}}$bingen, Germany}  and A.E.L. Dieperink}
\affiliation{Theory Group, Kernfysisch Versneller Instituut, University of Groningen,
Zernikelaan 25, NL-9747 AA Groningen, The Netherlands}
\begin{abstract}
The contribution from hyperons to the bulk viscosity of neutron star matter is calculated. Compared
to previous works we use for the weak interaction the one-pion
exchange model rather than a current-current interaction, and
include the neutral current $nn\leftrightarrow n\Lambda$ process.
Also the sensitivity to details of the equation of state is examined.
Compared to previous works we find that the contribution from hyperons to the
bulk viscosity is about two orders of magnitude smaller.
\end{abstract}
\maketitle
\section{Introduction}
\label{chap:hypvis;sec:int}
The bulk viscosity of matter in neutron stars has recently
received considerable attention in connection with damping of
neutron star pulsations and gravitational radiation driven
instabilities, especially in the damping of $r$-modes~\cite{And2001}.
If the $r$-modes are unstable, i.e. if the damping time-scales
due to viscous processes in neutron star matter are longer than
the gravitational radiation driving time-scale, a rapidly rotating
neutron star could emit a significant fraction of its rotational
energy and angular momentum as gravitational waves, which could be
detectable.
\\ \indent
It has been shown~\cite{Hae2000,Hae2001a,Hae2002,Jon2001,Lin2002}
that the bulk viscosity in a neutron star is caused by energy
dissipation associated with  nonequilibrium weak interaction
reactions in a pulsating dense matter. Strong interaction
processes do not play a role, because the strong interaction
equilibrium is reached so fast that these processes can be
considered to be in thermal equilibrium compared to the typical
pulsation time-scales of $10^{-4}-10^{-3} \ \textrm{s}$. The
relevant nonequilibrium   weak interaction reaction rates and the
magnitude of the bulk viscosity depend on the density and the
composition of neutron star matter.
\\ \indent
In a relatively low density neutron star composed mainly of
neutrons $n$ with admixtures of protons $p$, electrons $e$, and muons $\mu$,
the bulk viscosity is mainly determined by  the reactions of the nonequilibrium
modified Urca process,
\begin{eqnarray}
N \ + \ n \ \rightarrow \ N \ + \ p \ + \ l \ + \ \overline{\nu}_l,
\ \ N \ + \ p \ + l \ \rightarrow \ N \ + \ n + \ \nu _l,
\label{eq:modurca}
\end{eqnarray}
with $N=n, p$  and  $l=e, \mu$. The bulk viscosity was studied by
Sawyer~\cite{Saw1989} for $npe$ matter and by Haensel et
al.~\cite{Hae2001a} for $npe\mu$ matter. For the process in
Eq.~(\ref{eq:modurca}) the relaxation time,
i.e. the time it takes to restore equilibrium in case of a
perturbation, is strongly temperature dependent, namely $\tau ^{-1} \propto T^6$.
\\ \indent
At densities $n_B$ of a few times saturation density,
$n _0$ ($n _0 \approx 0.15 \ {\rm fm}^{-3}$), the direct Urca process,
\begin{eqnarray}
n \ \rightarrow \ p \ + \ l \ + \ \overline{\nu}_l, \ \ p \ + \ l \ \rightarrow \ n \ + \ \nu _l,
\label{eq:directurca}
\end{eqnarray}
may also be allowed depending on whether or not the proton fraction
exceeds the Urca limit of $x_p \approx 0.11$ \cite{Lat1991a}. The
contribution of this process to the bulk viscosity was computed
by Haensel and Schaeffer~\cite{Hae1992} for $npe$ matter and
by Haensel \textit{et al.}~\cite{Hae2000} for $npe\mu$ matter. Compared to
the modified Urca process a smaller number of particles is
involved 
which leads to a weaker temperature dependence, $\tau ^{-1}
\propto T^4$. As a consequence at typical neutron star
temperatures, $ T \sim 10^9-10^{10} \ \textrm{K}$, its contribution
to the bulk viscosity is typically $4-6$ orders of magnitude
larger than that from the modified Urca process. The largest difference compared
with that from the modified Urca process is reached at the low temperatures.
\\ \indent At about the same densities hyperons may appear in the neutron star
core, first the $\Sigma ^{-}$ and $\Lambda $ hyperons, followed by
$\Xi ^0$, $\Xi ^{-}$, and $\Sigma ^{+}$ at higher densities. Here
we will restrict ourselves to the $\Sigma ^{-}$ and $\Lambda$
hyperons. In addition to the semileptonic hyperon processes, weak
nonleptonic hyperon processes also occur, specifically the
processes
%
\begin{equation}
n \ + \ n \ \leftrightarrow \ p \ + \ \Sigma ^{-},
\label{nnpS}
\end{equation}
%
%
\begin{equation}
p \ + \ n \ \leftrightarrow \ p \ + \ \Lambda,
\label{pnpL}
\end{equation}
and 
\begin{equation}
n \ + \ n \ \leftrightarrow \ n \ +  \ \Lambda.
\label{nnnL}
\end{equation}
At low temperatures these processes contribute more efficiently to the
bulk viscosity than the direct Urca process
and the semileptonic hyperon ones, because they contain no neutrino
phase space factor; at typical neutron star temperatures,
$T< 10^{10} \ \textrm{K}$, the phase space of neutrinos is almost negligible
compared to that of baryons. Hence, for the weak nonleptonic
hyperon processes of Eqs.~(\ref{nnpS}-\ref{nnnL}) the temperature
dependence of the inverse relaxation time  is $\tau ^{-1} \propto
T^2$.
%
%
%
%
\\ \indent
Historically the first semi-quantitative calculation of bulk
viscosity in neutron matter was carried out by Jones \cite{Jon1971}. In
this and all later works the weak nonleptonic process was
calculated using a baryon current-current interaction, i.e. a
contact $W$ exchange.

More recently the contribution to the bulk
viscosity from the various  weak nonleptonic hyperon processes has
been reconsidered by several authors using a modern equation of state (EoS). Haensel et
al.~\cite{Hae2002} studied the bulk viscosity for
the $nnp\Sigma ^{-}$ process 
within the nonrelativistic limit, and they found the bulk
viscosity to be several orders of magnitude larger than that of
the direct and the modified Urca processes. Also
applying the contact interaction of $W$ exchange, Lindblom
and Owen~\cite{Lin2002} calculated the contribution  of the
$pnp\Lambda$ process in addition to that of the $nnp\Sigma ^{-}$
process.
\\  \indent One may question the validity of the $W$ exchange process.
First there is no
$W$ exchange contribution to the $nnn\Lambda$ process and
therefore it has not been considered in the above approaches~\cite{Jon1971,Hae2002,Lin2002}.
On
the other hand it is well known that in nuclear physics
experiments on weak $\Lambda$ decay in large hypernuclei the
$nnn\Lambda$ process is found equally important as the
$pnp\Lambda$ process. Secondly, Jones noticed already several orders of magnitude
difference between $\tau _{nnn\Lambda}$ and the theoretical $W$ exchange
based value of $\tau _{nnp\Sigma ^-}$~\cite{Jon2001}.
This is interesting, because the rate for the $nnn\Lambda$ process
must depend on a very wide class of possible weak hadronic processes.
Because most of these processes also contribute to $nnp\Sigma ^-$
process, the $\tau _{nnp\Sigma ^-}$ in reality
is probably of the same order of magnitude as $\tau _{nnn\Lambda}$.
\\ \indent However, in nuclear physics it is customary to model the weak nonleptonic
$NY \to NN$ process in terms of meson exchange (pion and kaon)
with one phenomenological weak $\pi NY$ or $KNN$ vertex. In this
approach one can describe both the  observed weak decay rate of
the  $\Lambda$ as well as branching ratios in hypernuclei quite well~\cite{Alb2002}.
Therefore, in the following we will  calculate the
bulk viscosity using the meson exchange picture to
describe the hyperon processes in Eqs.~(\ref{nnpS}-\ref{nnnL}). \\
\indent
In practice the rates also depend on the details of the equation
of state, e.g. the hyperon fractions as a function of the
density. A variety of model equations of state have been
constructed with widely varying properties during the last
decades, some of these based upon a microscopic free space $NN$ and
NY interactions, while others are phenomenological
parametrizations of the energy density to higher densities. In
particular
the relativistic mean field approaches, as used in Ref.~\cite{Hae2002},
and Ref.~\cite{Lin2002}, although microscopic in nature, do not start
from a realistic nucleon-nucleon interaction. 
Here we employ the effective EoS based on the work of
Balberg and Gal~\cite{Bal1997}. To get some idea about the
sensitivity  of the bulk viscosity to the details of the EoS we
consider two different parameter sets for the density dependence
of the multi-body potential energy.
\\ \indent
Superfluidity can also influence the bulk viscosity.
Below the critical temperature, superfluidity~\cite{Hae2002,Lin2002}
suppresses the reaction rates by roughly a factor $\exp(-\Delta /T)$
with gap energy $\Delta$. It generally leads to a smaller bulk viscosity.
Above the critical temperature, it has no influence on the reaction rates.
Many properties of superfluid matter such as the gaps and
the critical temperatures are still known with large uncertainties, e.g. $T_c \sim 10^{8}-10^{10} K$.
Hence, the bulk viscosity is only considered here for
nonsuperfluid matter.
For recent papers about the effects of superfluidity we refer
to Haensel \textit{et al.}~\cite{Hae2002} and Lindblom and Owen~\cite{Lin2002}.   \\ \indent
The main goal in this study is threefold (i) to compute the
viscosity using the  more realistic one-pion exchange (OPE)
instead of the contact $W$ exchange description in the
nonleptonic weak hyperon processes, (ii) to include the weak
neutral current  $nnn\Lambda$ process, and (iii) to
examine the sensitivity of the bulk viscosity to the EoS. The
starting point of our derivation of the bulk viscosity for the
$nnp\Sigma ^{-}$, $pnp\Lambda$, and $nnn\Lambda$ processes is the
finite temperature Green's function formalism in the quasi-particle
approximation (QPA). The coupling constants for these processes
using OPE can be verified from pionic hyperon decay experiments.
The application of these coupling constants gives the correct
order of magnitude for the rate of nonmesonic hyperon decay in
large hypernuclei. Covariant expressions for the matrix elements
are derived.
\\ \indent The plan of this paper is as follows. The EoS
is discussed in Section~\ref{chap:chap:vis;sec:EoS}. The bulk
viscosity for neutron star matter is discussed in
section~\ref{chap:vis;sec:gen}. Section~\ref{chap:vis;sec:colrate}
is devoted to the collision rate of the weak nonleptonic hyperon
process as in Eqs.~(\ref{nnpS}-\ref{nnnL}). The results are presented
and discussed in section~\ref{chap:vis;sec:resdis}. Finally the
conclusion is given in section~\ref{chap:vis;sec:con}.
\section{Equation of State}
\label{chap:chap:vis;sec:EoS}
\begin{figure}
\begin{center}
\includegraphics[width=0.9\textwidth] {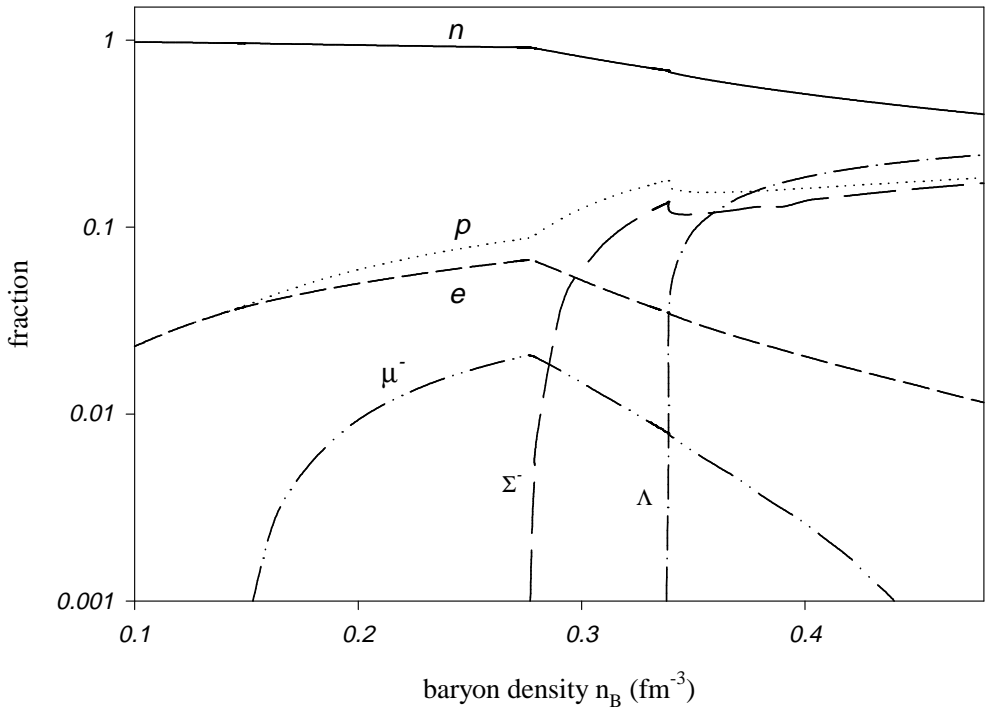}
\caption{The particle fraction as a function of the baryon number
density (Balberg and Gal~\cite{Bal1997} case 1 with $\gamma =\delta = 2$).
\label{fig:fractions_case1}}
\end{center}
\end{figure}
\begin{figure}
\begin{center}
\includegraphics[width=0.9\textwidth] {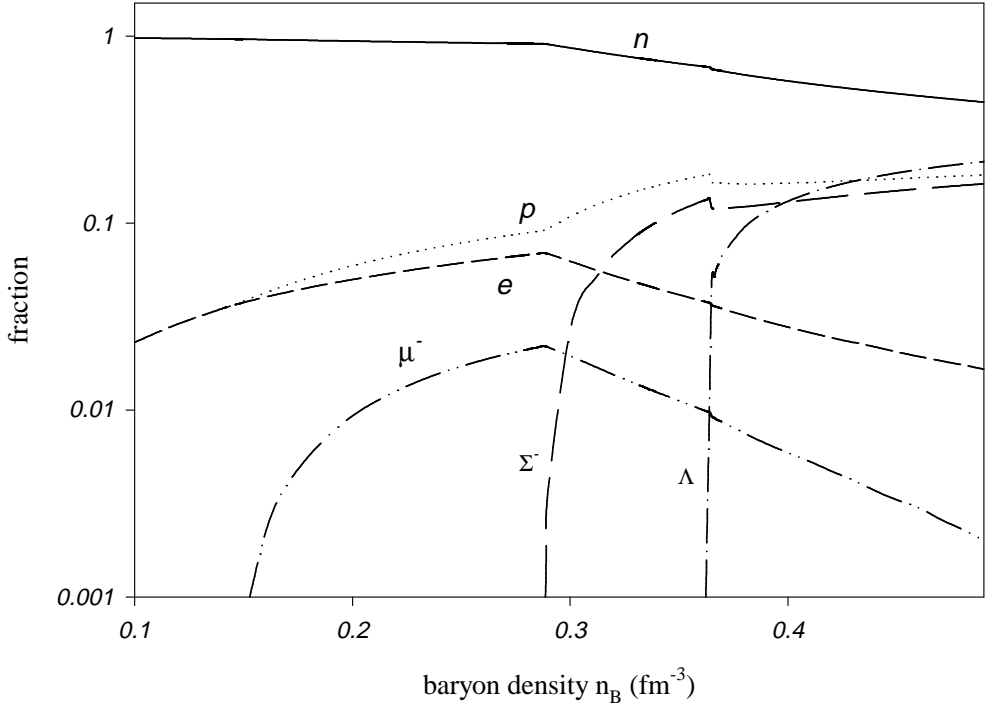}
\caption{The particle fraction as a function of the baryon number
density (Balberg and Gal~\cite{Bal1997} case 2 with $\gamma =\delta = 5/3$).
\label{fig:fractions_case2}}
\end{center}
\end{figure}
In this paper we use the equation of state constructed by Balberg
and Gal~\cite{Bal1997}. It is a generalization of the (npe) EoS of Lattimer and Swesty~\cite{Lat1991b}.
The energy density is parametrized in terms of the densities of the constituents.
The parameters for the nucleon-nucleon interactions are fitted to the binding energy, symmetry energy,
and incompressibility of saturated symmetric nuclear matter. The values of the parameters for hyperon-nucleon
and hyperon-hyperon interactions mainly rely on experimental data from hypernuclei.
The equilibrium fractions are obtained in the standard manner, by
assuming a given nucleon density  $n$ and solving the imposed
conditions of baryon conservation, charge neutrality and weak
equilibrium \begin{eqnarray}  n_B & = &  n_n \ + \ n_p \ + \ n_{\Sigma ^-} \ + \ n_{\Lambda} \label{eq:barcon}, \\
 n_p & = & n_e\ + \ n_{\Sigma ^-}, \\
\mu_p & = & \mu_n \ - \ \mu_e, \\
\mu_e & = & \mu_\mu, \\ \mu_{\Sigma ^-} & = & \mu_n \ + \ \mu_e, \\
\mu_\Lambda & = &
\mu_n. \label{eq:lambdaweak}
\end{eqnarray}
\\ \indent To obtain some idea about the sensitivity
of the  resulting bulk viscosity to details of the EoS we use
two parametrizations which differ in the density dependence of
the contributions of the nucleons ($\delta)$ and hyperons
($\gamma)$ to the energy density, specifically given by the
parameter sets $\gamma =\delta = 2$ and $\gamma =\delta = 5/3$,
respectively.
The difference between the two parametrizations of the equation of
state has no influence on the particle fractions in the nuclear
sector. For densities below the appearance density of the hyperons the
neutrons are abundantly present ($x_n > 0.9$). The leptons are
present because of charge neutrality to compensate the positively
charged protons, although they are expensive in terms of energy
density. The appearance of the hyperons create the possibility to
lower the neutron excess without lepton formation, whereas the
negatively charged hyperons allow charge neutrality to be
maintained within the baryon sector. Therefore, the appearance of the
hyperons is accompanied by a strong deleptonization.
\\ The first hyperon to appear is the $\Sigma ^{-}$, followed by the
lower-mass $\Lambda$. This can be explained by the higher chemical
potential of the $\Sigma ^{-}$ due to its negative charge, which
compensates the mass difference of about 80 MeV. However, the
growth of the $\Sigma ^{-}$ number density fraction is soon
hindered by charge dependent forces, which disfavor an excess of
$\Sigma ^{-}$'s over $\Sigma ^{+}$'s and a joint excess of $\Sigma
^{-}$'s and neutrons. The $\Lambda$ is not affected by
charge-dependent forces and its fraction continues to increase.
For the EoS with the smaller values for $\delta=\gamma= 5/3$ the onset of the various hyperons occur at  higher
densities and the corresponding hyperon fractions are smaller.
\section{The Bulk Viscosity}
\label{chap:vis;sec:gen}
In general bulk viscosity is a dissipative process in which the
compression is converted into heat. It is due to the instantaneous
difference between the local physical pressure $p$ and the
thermodynamic pressure $\tilde{p}$ $$ p \ - \ \tilde{p} \ =  \ -\zeta \ \vec{\nabla}.\vec{v},
$$ where $\vec{v}$ is the local velocity. In neutron star matter one is
interested in describing small deviations (oscillations) $\delta
x_i =  x_i  - \tilde{x}_i $ around the thermodynamic equilibrium,
characterized by the variables (densities)$\tilde{x}_i.$
\\ \indent We
will now turn to the calculation of $\zeta$ for the specific case of
a neutron star with hyperons. Since the three reactions~(\ref{nnpS}-\ref{nnnL})
involve neutrons the  fluctuation of the neutron fraction $x_n$
can be used as the primary parameter.
 For this situation a  general expression for the bulk
viscosity  has been derived in Ref.~\cite{Lin2002}
\begin{eqnarray}
\zeta \ = \ \frac{-n_B \ \tau}{1 \ - \ i \ \omega \ \tau} \bigg(\frac{\partial
p}{\partial x_n}\bigg)_{n_B} \frac{d\tilde{x}_n}{dn_B},
\end{eqnarray}
with $\tau$ the relaxation time, $\omega$ the pulsation frequency
of the neutron star, $p$ the pressure, $n_B$ the baryon density.
\\ \indent
 We restrict ourselves to the
densities, at which nucleons, $\Sigma ^{-}$'s, and $\Lambda$'s
are present. Defining the chemical imbalance as
\begin{eqnarray}
\chi \ = \ 2 \ \mu _n \ - \ \mu _p \ - \ \mu _{\Sigma ^-} \ = \ \mu _n \ - \ \mu
_{\Lambda},
\end{eqnarray}
and assuming that in first order 
the difference $\Delta \Gamma_a$ between the rates for the various direct reactions $\Gamma _a$ and inverse
reactions $\bar{\Gamma} _a$ in Eqs.~(\ref{nnpS}-\ref{nnnL})
is proportional to $\chi$
\begin{eqnarray}
\Delta \Gamma \ = \ \sum_a \ \Delta \Gamma_a \ = \ \lambda \ \chi,
\label{eq:Gammalambda}
\end{eqnarray}
with $\lambda$ determining the viscosity. Hence,
 the relaxation time $\tau$ can be expressed in terms
of the variation of the imbalance with the neutron density
fluctuation
 \begin{eqnarray} \tau \ = \ \frac{n_B \ \chi}{ \Delta \Gamma \
(d \chi  / d x _{n})} \approx \frac{n_B}{\lambda \
(d \chi  /
d x _{n})}. \label{eq:reltimelambdasigma}
\end{eqnarray}
Thus the main task is to evaluate the $d \chi/ d x_n.$
We wish to
include several weak nonleptonic hyperon reactions. To do this in
a proper way one has to take into account that  the four relevant
baryon number densities, $x_n, x_p, x_\Lambda$, and $x_\Sigma,$ are
related to each other by the following three constraints: baryon
conservation, charge neutrality, and chemical equilibrium for
strong processes and in particular for the reaction $ n+ \Lambda
\leftrightarrow p+ \Sigma ^- $.
By using these conditions one can express
$\chi$ in terms of the fractions $x_i$ and
assuming that all leptonic reaction rates are much smaller than
those which produce the hyperon bulk viscosity, one obtains
in the density region with $np\Sigma^-$ matter ~\cite{Lin2002}
\begin{eqnarray}
\frac{d \chi}{d x_n} & = & 2 \ \alpha _{nn} \
- \ ( \ \alpha _{pn} \ + \ \alpha _{\Sigma ^{-}n} \ + \ \alpha _{np} \ + \ \alpha _{n\Sigma ^{-}} \ )
\nonumber\\  &&
+ \ (1/2) \ (\alpha _{pp} \ + \ \alpha _{\Sigma ^{-}p} \ + \ \alpha _{p \Sigma ^{-}} \ + \ \alpha _{\Sigma ^{-}\Sigma ^{-}}),
\end{eqnarray}
and in the region with $np\Sigma ^- \Lambda $ matter
\begin{eqnarray}
\frac{d \chi}{d x_n} & = & \alpha _{nn} \
+ \ \frac{( \ \beta _n \ - \ \beta _{\Lambda} \ ) ( \ \alpha _{np} \ - \ \alpha _{\Lambda p} \
+ \  \alpha _{n \Sigma ^{-}} \ - \ \alpha _{\Lambda \Sigma ^{-}} )}
{ 2 \ \beta _{\Lambda} \ - \ \beta _p \ - \ \beta _{\Sigma ^{-}}}
\nonumber\\  &&
- \ \alpha _{\Lambda n} \ - \frac{( \ 2 \ \beta _n \ - \ \beta _p \ - \ \beta _{\Sigma ^{-}} \ )
( \ \alpha _{n \Lambda} \ - \ \alpha _{\Lambda \Lambda} \ )}
{2 \ \beta _{\Lambda} \ - \ \beta _p \ - \ \beta _{\Sigma ^{-}}},
\end{eqnarray}
where the $\alpha$'s correspond to the partial derivatives of the chemical potentials with
respect to the various fractions
\begin{eqnarray}
\alpha _{ij} \ = \ (\frac{\partial \mu _i}{\partial x_j})_{n_k,k\not{=}j},
\end{eqnarray}
and
\begin{eqnarray}
\beta _i \ = \ \alpha _{ni} \ + \ \alpha _{\Lambda i} \ - \ \alpha _{p i} \ - \ \alpha _{\Sigma ^{-}i},
\end{eqnarray}
where $i$ and $j$ stand for $n$, $p$, ${\Sigma}^{-}$, and
$\Lambda$. These quantities are obtained from the EoS.
\\ \indent The real part of the hyperon bulk viscosity
\begin{eqnarray}
\Re e \ \zeta  & = & \frac{- n_B \ \tau }{1 \ + \ \omega ^2 \ \tau ^2}
\Big( \frac{\partial P}{\partial x _{n}} \Big) \ \frac{d x _{n}}{d n_B},
\end{eqnarray}
where $\tau$ is given in Eq.~(\ref{eq:reltimelambdasigma}). The
core of the neutron star is assumed to pulsate with a frequency of about
$\omega \sim 10^{3}-10^{4} \ \textrm{s}^{-1}$. In the high frequency limit $1
<< \omega \tau$ the bulk viscosity $\zeta$ is proportional to the
inverse of the relaxation time, $\zeta \propto \tau ^{-1} \propto
\lambda$, whereas in the low frequency limit $1 >> \omega \tau$
the bulk viscosity is proportional to the relaxation time, $\zeta
\propto \tau \propto \lambda ^{-1}$. In the following section the
difference in the rates given by $\Delta \Gamma$ in
Eq.~(\ref{eq:Gammalambda}) in first order in $\chi$ is derived, from which $\lambda =
\Delta \Gamma / \chi $ can be obtained.
\section{Collision rate}
\label{chap:vis;sec:colrate}
In this section  the various rates $\Delta \Gamma _{\sss N1N2N'Y'}$ for
the processes  $N_1 + N_2 \leftrightarrow N' + Y'$ are considered,
where the $N$ is a nucleon and the $Y$ is a hyperon.
First we consider the one-pion exchange in Born approximation.
For completeness the rates
using the contact $W$ exchange interaction~\cite{Lin2002}
are given  in section~\ref{chap,sec,subsec:vis,colrate,W}.
%
\subsection{One-pion exchange}
\label{chap,sec,subsec:vis,colrate,OPE}
\begin{figure}[!h]
\begin{eqnarray}
\includegraphics[width=0.3\textwidth] {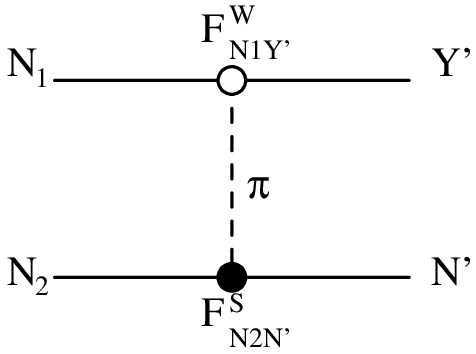} \ \ \ \ \ \ & \ \ \ \ \ \
\includegraphics[width=0.3\textwidth] {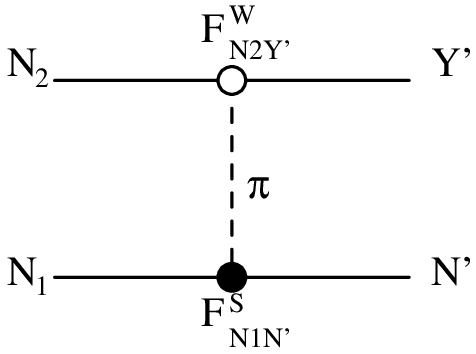}  \nonumber
\end{eqnarray}
\caption{The pion exchange diagrams in Born approximation
for the weak nonleptonic hyperon processes
with only one hyperon involved. \label{diany}}
\end{figure}
%
%
In order to obtain
$\Delta \Gamma _{\sss N1N2N'Y'}$, first the rate $\Gamma _{\sss N1N2 \rightarrow N'Y'}$ is considered.
The relevant free space diagrams are shown
in Fig.~\ref{diany}. The strangeness changing weak vertex is given by
\begin{eqnarray}
F ^w_{\sss NY'} \ = \ G_F \ m_{\pi}^2 \ ( \ \bar{A}_{\sss NY'} \
+ \ \bar{B}_{\sss NY'} \ \gamma _5 \ ),
\label{strongcoup}
\end{eqnarray}
for which the constants $\bar{A}$ and $\bar{B}$ determine the strengths of the
parity violating and parity conserving  $Y \rightarrow N + \pi$
amplitudes, respectively, and are specified
in section~\ref{chap:vis;sec:resdis}.
%
The strong vertex is
\begin{eqnarray}
F ^s_{\sss N N'} \ = \ g_{\sss N N'} \ \gamma _5
\end{eqnarray}

\begin{figure}[!tbp]
\begin{eqnarray}
\includegraphics[width=0.3\textwidth] {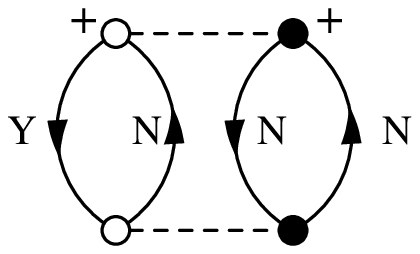} \ \ \ \ \ \ & \ \ \ \ \ \
\includegraphics[width=0.3\textwidth] {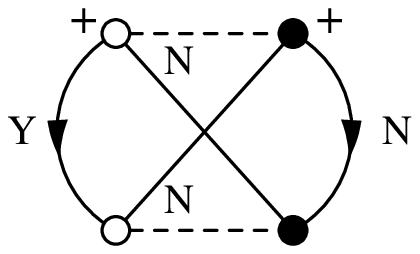}  \nonumber
\end{eqnarray}
\caption{The two types of closed diagrams at the two-loop level.}
\label{twoloop hyperon}
\end{figure}

 \indent To compute the rates in the medium one needs to account
for Pauli blocking; it is convenient to use the optical theorem
to convert the free space diagrams of Fig.~\ref{diany} to the
closed diagrams of Fig.~\ref{twoloop hyperon}.
%
%
Using finite temperature Green functions in the QPA, one can express the collision rate as
\begin{eqnarray}
\Gamma_{\sss N1N2 \rightarrow N'Y'} & = &
\frac{1}{S} \ \int \ \frac{d^4p_{\sss N1}}{(2 \ \pi)^4} \ \frac{d^4p_{\sss N2}}{(2 \ \pi)^4} \
\frac{d^4p_{\sss N'}}{(2 \ \pi)^4} \ \frac{d^4p_{\sss Y'}}{(2 \ \pi)^4} \nonumber\\ &&
Z \ (2 \pi)^4 \
\delta ^4(p_{\sss N1}+p_{\sss N2}-p_{\sss Y'}-p_{\sss N'}).
\label{eq:colrate}
\end{eqnarray}
In Eq.~(\ref{eq:colrate}) one has
\begin{eqnarray}
Z=Z^a+Z^b,
\end{eqnarray}
with
\begin{eqnarray}
Z^a & = & \Tr \Bigg[ \ G_{\sss N1}^{- +}(p_{\sss N1}) \ F ^s_{\sss N1N'} \ G_{\sss N'}^{+
-}(p_{\sss N'}) \ F ^{s^{\dagger}}_{\sss N1N'} \Bigg] \ \nonumber\\ && \Tr \Bigg[ \
G_{\sss N2}^{- +}(p_{\sss N2}) \ F ^w_{\sss N2Y'}  \ G_{\sss Y'}^{+ -}(p_{\sss Y'}) \
F^{w^{\dagger}}_{\sss N2Y'} \ \Bigg] \ D_{\pi}(k_1^2)^2
\nonumber\\ && + \ \{ N_1 \ \leftrightarrow \ N_2 \}
\end{eqnarray}
with $D_\pi(k^2)= \frac{1}{\vec{k}^2+m_\pi^2}$, and $\vec{k}_i=\vec{p}_{\sss Ni}-\vec{p}_{\sss N'}$ (for $i=1,2$), corresponding to the
diagram on the left in Fig.~\ref{twoloop hyperon}  and
\begin{eqnarray}
Z^b & = & \Tr \Bigg[ \ G_{\sss N1}^{- +}(p_{\sss N1}) \ F ^s_{\sss N1N'} \ G_{\sss N'}^{+
-}(p_{\sss N'}) \ F ^{s^{\dagger}}_{\sss N2N'} \ G_{\sss N2}^{- +}(p_{\sss N2}) \ F ^w_{\sss N2Y'} \
G_{\sss Y'}^{+ -}(p_{\sss Y'}) \ F ^{w^{\dagger}}_{\sss N1Y'} \ \Bigg] \nonumber\\  & &
D_{\pi}(k_1^2) \  D_{\pi}(k_2^2) + \{ N_1 \leftrightarrow N_2 \}
\end{eqnarray}
corresponding to that of
the right in Fig.~\ref{twoloop hyperon}. The symmetry factor $S$
is for the $pnp\Lambda$ process $S=1$ and for the $nnp\Sigma$ and
$nnn\Lambda$ processes $S=2$. After evaluation of the $p^0$
integrals in the QPA, one obtains
\begin{eqnarray}
\Gamma _{\sss N1N2 \rightarrow N'Y'} & = & \int \ \frac{d^3p_{\sss N1}}{2 \
m_{\sss N1} \ (2 \ \pi)^3} \ \frac{d^3p_{\sss N2}}{2 \ m_{\sss N2} \ (2 \ \pi)^3}
\frac{d^3p_{\sss N'}}{2 \ m_{\sss N'} \ (2 \ \pi)^3} \ \frac{d^3p_{\sss Y'}}{2 \ m_{\sss Y'} \ (2 \
\pi)^3} \nonumber\\ && \frac{1}{S} \ |M_{\sss N1N2N'Y}|^2 \ f_{\sss N1} \ f_{\sss N2}  \
(1-f_{\sss N'}) \ (1-f_{\sss Y'}) \nonumber\\ && (2 \ \pi)^4 \ \delta(E _{\sss N1} + E_{\sss N2}  -
E _{\sss N'} -  E _{\sss Y'}  )   \nonumber\\ && \delta
^3(p_{\sss N1}  +  p_{\sss N2}  -  p_{\sss N'}  -  p_{\sss Y'}  ),   \label{rateNNNY}
\end{eqnarray}
where $p_i$ and $E_i$ are the particle momentum and energy, respectively,
and $f_i=\{ 1 + \exp[(\epsilon _i - \mu _i)/T]\}^{-1}$ is the Fermi-Dirac
distribution function.
The matrix element is
%
%
\begin{eqnarray}
M_{\sss N1N2N'Y'} & = & [ \ \bar{u}_{\sss N'} \ F ^s_{\sss N1N'} \ u_{\sss N1} \ \bar{u}_{\sss Y'} \
F ^w_{\sss N2Y'} \ \bar{u}_{\sss N2} \ D_{\pi}(k_1^2) \nonumber\\  &&
- \ \bar{u}_{\sss N'} \ F ^s_{\sss N2N'} \ u_{\sss N2} \ \bar{u}_{\sss Y'} \ F ^w_{\sss N1Y'}
\ \bar{u}_{\sss N1} \ D_{\pi}(k_2^2) \ ].
\label{eq:MN1N2N'Y'}
\end{eqnarray}
The evaluation of $\Gamma _{\sss N1N2 \rightarrow N'Y'} $ with the standard
technique~\cite{Sha1983}
takes advantage of the strong degeneracy
of the participating particles in neutron star matter.
The multidimensional integral is decomposed into an
energy and an angular integration. All momenta $p_i$
are placed on the appropriate Fermi spheres
wherever possible. Furthermore, we introduce the dimensionless
quantities
\begin{eqnarray}
y_i \ = \ \frac{\epsilon _i \ - \ \mu _i}{T}; \ \xi = \frac{\chi}{T}.
\end{eqnarray}
The rate in Eq.(\ref{rateNNNY}) can be factorized in the form
\begin{equation}\Gamma _{\sss N1 N2
\rightarrow N'Y'} \ =  \ \Gamma ^{(0)}_{\sss N1 N2 N'
Y'} \ I(\xi),
\end{equation} with
\begin{eqnarray}
I(\xi) \ = \ \Bigg[ \ \prod _{i=1}^4 \ \int _{-\infty}^{\infty} \ dy_i \ f(y_i) \
\delta(\sum _{i=1}^4 y_i + \xi) \label{eint} \ =  \
 \frac{e^{\xi}}{e^{\xi} \ - \ 1} \frac{4 \ \pi ^2 \ \xi \ + \ \xi
^3}{6},
\end{eqnarray}
and
\begin{eqnarray}
\Gamma _{\sss N1N2N'Y'}^{(0)} & = & \frac{p_{F{\sss N1}} \ T^3}{8 \ (2 \ \pi)^6}
\int \ d\theta _{\sss N'} \ d\theta _{\sss Y'} \ \frac{\Theta(1 \ -\ |C(  \theta _{\sss N'},
\theta _{\sss Y'}  ) |)}{\sqrt{1 \ - \ C^2(  \theta _{\sss N'}, \theta _{\sss Y'}  )}}
\nonumber\\ && \frac{1}{S} \ \sum _{spin} \ |M _{\sss N1N2N'Y'}|^2,
\end{eqnarray}
where $p_{F{\sss Ni}}$ is the Fermi momentum of baryon $N_i$, $\theta _{\sss N'}$ is the angle between $\vec{p}_{\sss N1}$
and $\vec{p}_{\sss N'}$,
$\theta _{\sss Y'}$ is the angle between $\vec{p}_{\sss N1}$ and $\vec{p}_{\sss Y'}$, and
\begin{eqnarray}
C(\theta _{\sss N'}, \theta _{\sss Y'}) & = &
(- p^2_{F{\sss N1}} \ - \ p^2_{F{\sss N'}} \ - \ p^2_{F{\sss Y'}} \
+ \ 2 \ p_{F{\sss N1}} \ p_{F{\sss N'}} \ \cos(\theta _{\sss N'})  \nonumber\\ &&
+ \ 2 \ p_{F{\sss N1}} \ p_{F{\sss Y'}} \ \cos(\theta _{\sss Y'}) \
- \ 2 \ p^2_{F{\sss N'}} \ p^2_{F{\sss Y'}} \ \cos(\theta _{\sss N'}) \ \cos(\theta _{\sss Y'}) \nonumber\\ &&
+ \ p^2_{F{\sss N2}})
/(2 \  p^2_{F{\sss N'}} \ \sin(\theta _{\sss N'}) \ \sin(\theta _{\sss Y'})).
\end{eqnarray}
%
%
%
The conditions  in section~\ref{chap:chap:vis;sec:EoS},
Eqs.~(\ref{eq:barcon}-\ref{eq:lambdaweak}), apply in thermodynamic equilibrium. However, small
deviations from thermodynamic equilibrium occur just after a
neutron star is born. For the bulk viscosity the small deviations from the weak chemical
equilibrium are important. For this purpose  we need to consider
$\Delta \Gamma _{\sss N1N2N'Y'}=\Gamma _{\sss N1N2 \rightarrow N'Y'} - \Gamma _{\sss N' Y \rightarrow N1 N2}$,
where the inverse rate is $\Gamma _{\sss N' Y \rightarrow N1 N2}~=~\Gamma ^{(0)}_{\sss N1 N2 N'
Y'}~I(- \xi)$. Therefore, in the linear approximation, valid
for small deviations $| \xi |=| \chi | /T << 1$
one has  \begin{eqnarray}
\Delta \Gamma _{\sss N1N2N'Y'}=\Gamma ^{(0)}_{\sss N1 N2 N'
Y'} \Delta I, \end{eqnarray}
with
\begin{eqnarray}
\Delta I  = I(\xi) - I(- \xi) = \frac{4 \pi ^2}{3} \xi.
\end{eqnarray}
The $\Delta I$ is the same for the weak nonleptonic hyperon processes of Eqs.~(\ref{nnpS}-\ref{nnnL}).
%
%
The total rate considered for the hyperon bulk viscosity is 
\begin{eqnarray}
\Delta \Gamma \ = \ \Delta \Gamma _{pnp\Lambda} \ + \ \Delta \Gamma _{nnn\Lambda} \
 \ + \ 2 \ \Delta \Gamma _{nnp\Sigma ^{-}}.
\end{eqnarray}
This result is inserted into Eq.~(\ref{eq:reltimelambdasigma}) to
obtain the relaxation time.

%
%
\subsection{Contact $W$ exchange interaction}
\label{chap,sec,subsec:vis,colrate,W}
\begin{figure}[!h]
\begin{eqnarray}
\includegraphics[width=0.3\textwidth] {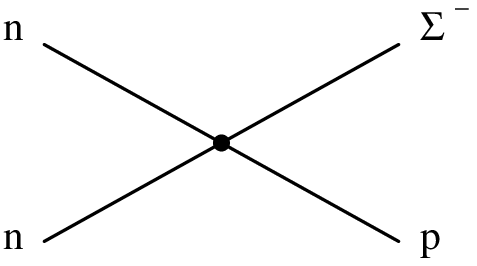} \ \ \ \ \ \ & \ \ \ \ \ \
\includegraphics[width=0.3\textwidth] {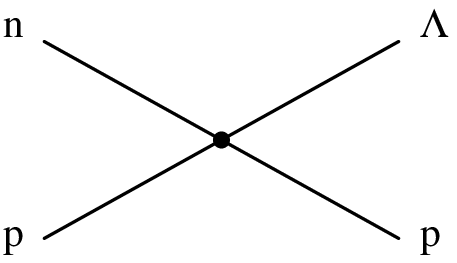}  \nonumber
\end{eqnarray}
\caption{The diagram of the contact $W$ exchange interaction
for the $nnp\Sigma ^{-}$ process
and the $pnp\Lambda$ process \label{diannpsigma}.}
\end{figure}
%
%
In order to illustrate the difference between the pion exchange
approach and the contact
$W$ exchange picture, 
we also consider the latter.
The rates for $W$ exchange have  been calculated before by
Haensel {\it et al.}~\cite{Hae2002} and by Lindblom and
Owen~\cite{Lin2002}, however in the angle-averaged approximation.
Performing an analogous derivation as in the previous section, one
obtains the expression in Eq.~(\ref{eq:reltimelambdasigma}). For
convenience of the reader here we give the simpler matrix elements
in the nonrelativistic limit (for the general expression for the
matrix elements we refer to
Ref.~\cite{Lin2002})
%

%
\begin{eqnarray}
|M_{nnp{\Sigma}^-}|^2 \ = \ 8 \ G_F^2 \ \sin^2(2 \ \theta _C)  \
m_n^2 \ m_p \  m_{{\Sigma}^-} \ (\ 1 \ + \ 3 \ c_A^{np} \ c_A^{n \Sigma ^-})^2
\label{eq:nnpSigmaW}
\end{eqnarray}
and
\begin{eqnarray}
|M_{pnp\Lambda}|^2 \ =  \
8 \ G_F^2  \ \sin^2(2 \ \theta _C) \ m_n \ m_p^2 \ m_{\Lambda} \ ( \ 1 \ + \ 3 \ |c_A^{np}|^2 \ |c_A^{p\Lambda}|^2 \ ),
\label{eq:pnpLambdaW}
\end{eqnarray}
respectively. Note that no meson propagators $D(k^2)$ are present  in Eqs.~(\ref{eq:nnpSigmaW})~and~(\ref{eq:pnpLambdaW}),
i.e. it has a different density dependence in the medium than the OPE matrix elements.
The $nnn\Lambda$ process has no simple $W$
exchange contribution. The results are shown in the following section.
\section{Results and Discussion}
\label{chap:vis;sec:resdis}
Some remarks about the differences between the meson exchange and
the $W$ exchange picture are appropriate. First in the OPE (in
which phenomenological input is used  for both the weak and strong
$\pi NN$ couplings) the pion exchange leads to a finite range
interaction and the pseudo scalar $\pi NN$ coupling to the
presence of momentum dependent terms, $p_i/m$. These effects are
absent in the simple $W$  point coupling. Hence, in addition to a
different overall strength one also expects to find a different
density dependence of the matrix elements for the two processes in
the medium. Secondly, the neutral current process, such as
$nnn\Lambda$ process, cannot be described in the simple $W$
exchange approach.

The EoS applied for the following figures is given by Balberg and
Gal~\cite{Bal1997} with parametrization $\gamma=\delta=2$ (case 1)
and also $\gamma=\delta=5/3$ (case 2). In the calculation for the
case of pion exchange we use the following values for the strong
and weak vertices: $g_{nn}^2/(4 \pi)=14.2$ and $g_{np}=\sqrt{2}
g_{nn}$. The values $\bar{A}_{n\Sigma ^-}=1.93$, $\bar{A}_{n
\Lambda}=-1.07$, $\bar{A}_{p \Lambda}=1.46$, $\bar{B}_{n \Sigma
^-}=-0.63$, $\bar{B}_{n \Lambda}=-7.19$, and $\bar{B}_{p
\Lambda}=9.95$ are based upon the experimental data on pionic
decay of hyperons~\cite{Sca2001}~\footnote{ $\bar{A}_{NY'} =
A_{NY'}/(G_F m_{\pi}^2)$ and $\bar{B}_{NY'} = B_{NY'}/(G_F
m_{\pi}^2)$ in Ref.~\cite{Sca2001} }. In the $W$ exchange picture
the standard values of the weak coupling constants are
${c_A^{np}}=-g_A=-1.26$, ${c_A^{p\Lambda}}=-0.72$, and
${c_A^{n{\Sigma}^-}}=0.34$ and $G_F$ is the Fermi weak coupling
constant. These values are also used by Lindblom and
Owen~\cite{Lin2002}, while Haensel \textit{et al.}~\cite{Hae2002}
treats the  interaction matrix as a free parameter.
Compared the work of
Lindblom and Owen~\cite{Lin2002} and Haensel \textit{et al.}~\cite{Hae2002} effects
of superfluidity are not taken into account and there is no angle averaging in this work.
\begin{figure}
\begin{center}
\includegraphics[width=0.8\textwidth] {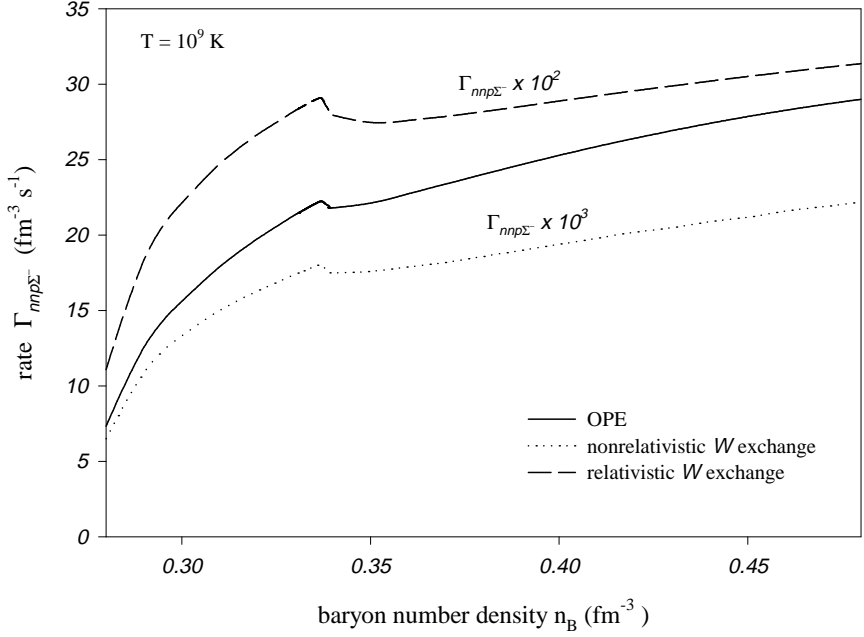}
\caption{The equilibrium rate $\Gamma$ for
$n + n \rightarrow p + \Sigma ^-$ as a function of the baryon number
density at $T = 10^9 \ \textrm{K}$.
\label{fig:ratennpSigma}}
\end{center}
\end{figure}
\\ \indent In  Fig.~\ref{fig:ratennpSigma} we compare the rates for the $n+n
\to p+\Sigma^- $ process for the $W$ exchange and the pion
exchange as a function of density for case 1 of the EoS. Concerning the $W$ exchange, we
note that the nonrelativistic result is a poor approximation to
the relativistic result since there is an almost complete
cancellation between the leading order vector and axial-vector
contributions, $(1+3 c_A^{np} c_A^{n{\Sigma}^-})^2 \approx 1/16$.
The overall rate calculated with OPE is almost two orders of
magnitude larger than the  one corresponding to relativistic $W$
exchange. The difference can be attributed to larger values of the
couplings, $\bar{A}_{NY'},\bar{B}_{NY'}, \ \textrm{and} \ g_{NN'}
> 1$, in the OPE case. At low momentum transfer (low density), this effect would
be compensated for by $p$-wave character of the $\pi NN$ coupling,
$\vec{\sigma}.\vec{k}$. However, for  momenta relevant in neutron
stars this does not happen. In addition, a stronger density
dependence is observed, which is related to finite range pion
exchange in Eq.~(\ref{eq:MN1N2N'Y'}).


Because of the rapid increase of the $\Lambda$ fraction after its
appearance, the proton and $\Sigma ^-$ fraction  drop abruptly and
also their Fermi momenta. Therefore, the kink in the rate can be
attributed to the appearance of the $\Lambda$ at $n_B \approx 0.33
\ {\rm fm}^{-3}$.
\begin{figure}
\begin{center}
\includegraphics[width=0.8\textwidth] {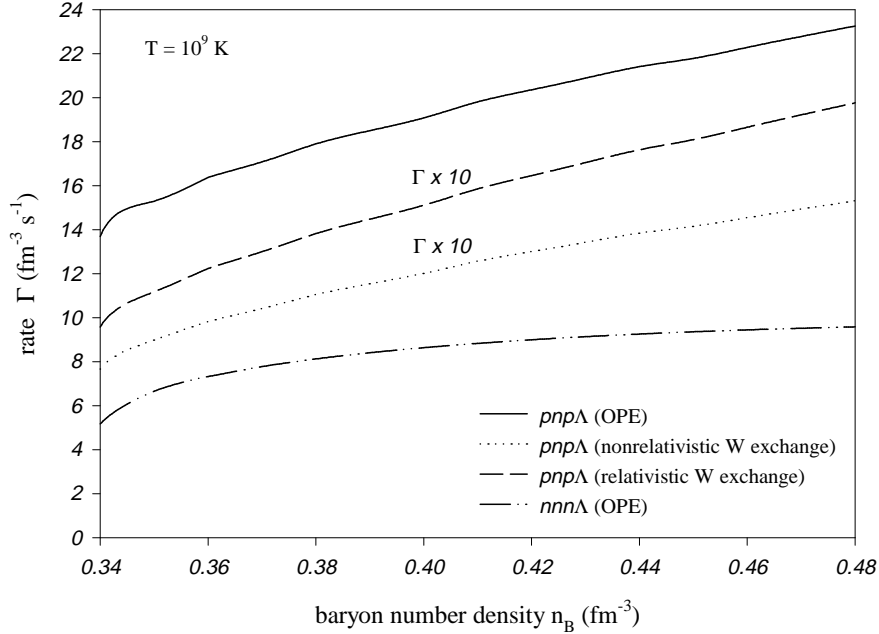}
\caption{The equilibrium rate $\Gamma$ for
$p + n \rightarrow p + \Lambda$ and $n + n \rightarrow n + \Lambda$
as a function of the baryon number
density at $T = 10^9 \ \textrm{K}$.
\label{fig:ratepnpLambda}}
\end{center}
\end{figure}
\\ \indent  The rates for $pnp\Lambda$ process are shown in Fig.~\ref{fig:ratepnpLambda}.
As mentioned above the OPE shows a stronger density dependence
because of the presence of the meson propagator. Although at
baryon densities where the hyperons are present  a partial
cancellation between the direct and exchange OPE matrix elements
occurs, the overall OPE rate is about one order of magnitude larger
than the one obtained with $W$ exchange, which again can be
attributed to the larger values of the meson couplings.

As mentioned above the neutral current $nnn\Lambda$ process does
not receive a contribution from $W$ exchange. In
the OPE approach the magnitude of this rate
is a factor 2-3 smaller than that of the $pnp\Lambda$ process. We
note that there exists  experimental information on the ratio of
the $nnn\Lambda$ and $pnp\Lambda$ processes in the weak decay of
hypernuclei which suggests that the simple OPE mechanism is not
sufficiently accurate \cite{Alb2002}. The inclusion of other
mesons, such as kaons, could improve the situation.
%
\begin{figure}
\begin{center}
\includegraphics[width=0.8\textwidth]{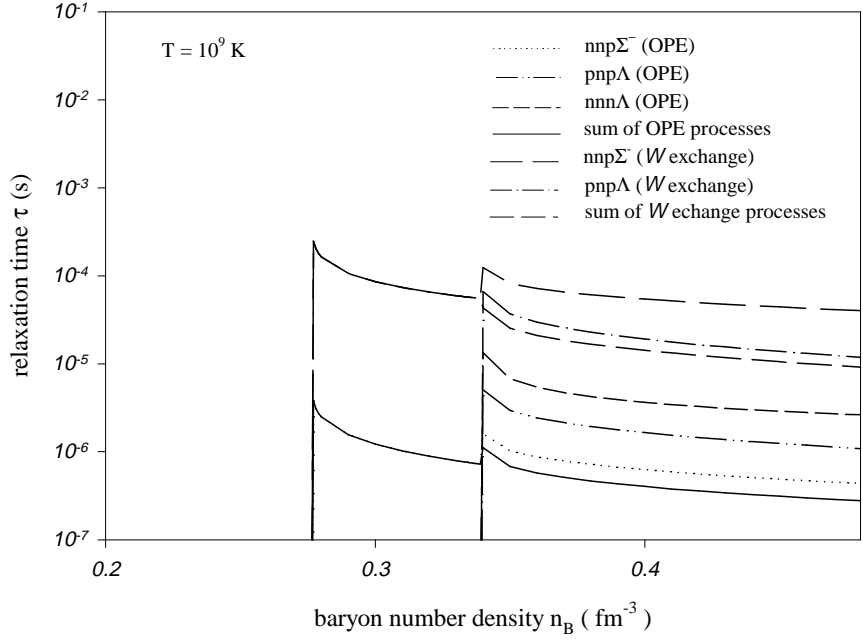}
\caption{The relaxation time $\tau$ as a function of the baryon number
density at $T = 10^9 \ \textrm{K}$.
\label{fig:relaxation_time}}
\end{center}
\end{figure}
\\ \indent  The relaxation time is shown
as a function of the baryon density in
Fig.~\ref{fig:relaxation_time}. One sees that the relaxation time
in the OPE picture is about two orders of magnitude smaller than
in the $W$ picture.
In the latter case after the appearance of $\Sigma ^-$, the
relaxation time is determined by the $nnp\Sigma ^-$ process; after
the occurrence of the $\Lambda$, the $pnp\Lambda$ process takes
over and it dominates the relaxation time. Therefore, a drop in
the relaxation time occurs at the appearance density of the
$\Lambda$.
 In the OPE picture after the
appearance of the $\Lambda$, the $nnp\Sigma ^-$ process remains
the most important process for the relaxation time. Therefore at
the appearance density of the $\Lambda$ a small abrupt increase in
the relaxation time occurs. \\ \indent
%
%
\begin{figure}
\begin{center}
\includegraphics[width=0.8\textwidth] {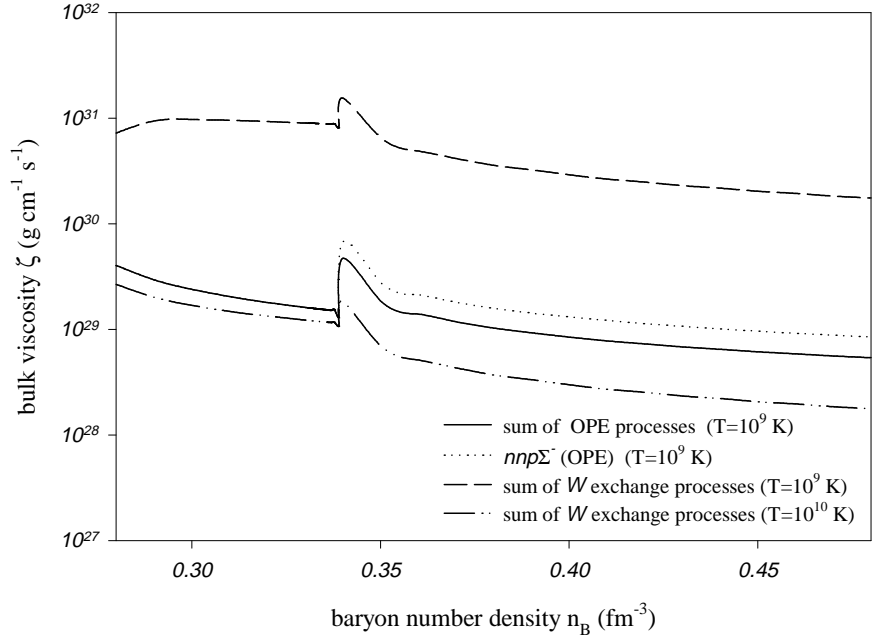}
\caption{The bulk viscosity $\zeta$ as a function of the baryon number
density.
\label{fig:bulk_viscosity}}
\end{center}
\end{figure}
To compute the viscosity one needs to assume a value for the
frequency;
 typical values of the frequency of the pulsations are $\omega
= 10^3-10^4 \ s^{-1}$. We have used $\omega=10^4 \ s^{-1}$ in
Fig.~\ref{fig:bulk_viscosity}. For the OPE picture, the high
frequency limit is applicable. Whereas for $W$ exchange at
$T=10^9 \ {\rm K}$ and at low density we are even in the low frequency
limit. The bulk viscosity in the OPE picture is about 1-2 orders of
magnitude smaller than that in the $W$ exchange picture.
\begin{figure}
\begin{center}
\includegraphics[width=0.99\textwidth] {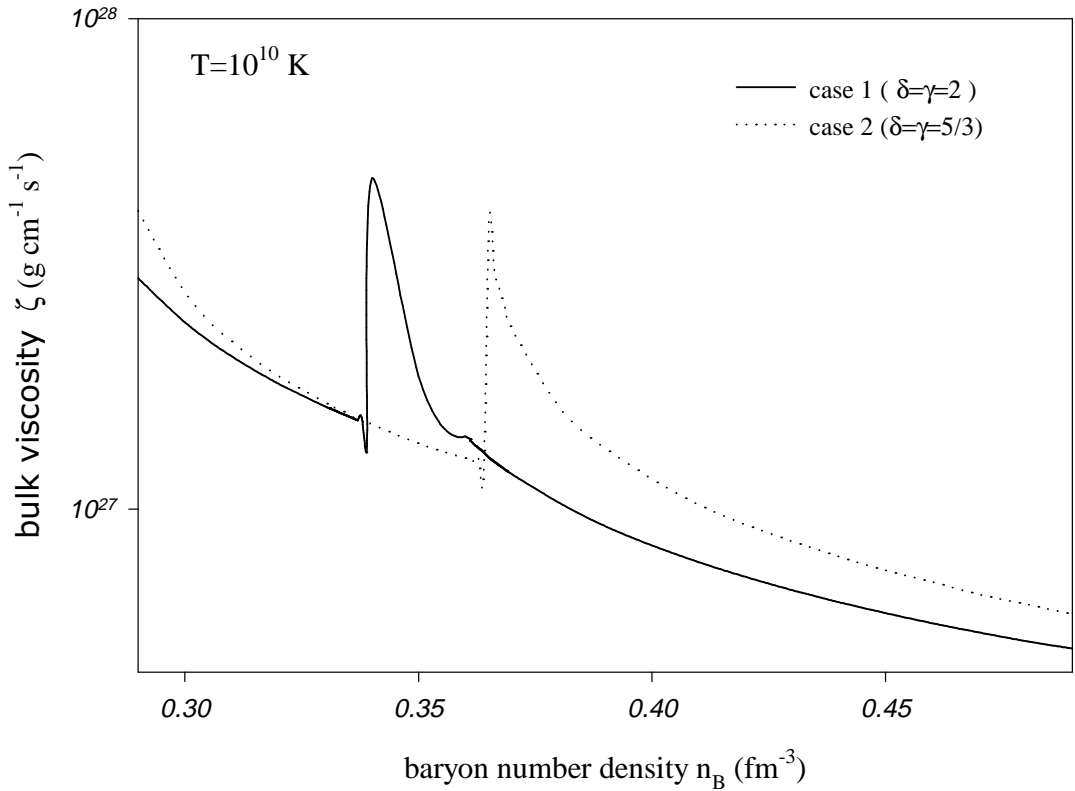}
\caption{The bulk viscosity $\zeta$ as a function of the baryon number
density for different equations of state at $T = 10^{10} \ K$.
\label{fig:bulk_viscosity_eos}}
\end{center}
\end{figure}
\\ \indent To investigate the sensitivity of the bulk viscosity  to details of
the EoS we also considered different values of the parameters. As can be seen in
Fig.~\ref{fig:bulk_viscosity_eos} the bulk viscosity is rather insensitive
to the details of the EoS except with respect to the appearance of the $\Lambda$.
\section{Conclusion}
\label{chap:vis;sec:con}
In this paper the bulk viscosity due to weak nonleptonic hyperon
processes has been studied. This viscosity is relevant in
connection with damping of neutron star pulsations, especially in
the damping of r-modes. In particular we considered pion exchange,
which is based upon empirical input, in Born approximation to
describe the weak nonleptonic hyperon processes instead of the $W$
exchange~\cite{Hae2002,Jon2001,Lin2002}. The conclusions are: i)
The bulk viscosity in the OPE picture is about 1-2 orders of magnitude
smaller than that in the $W$ exchange picture. ii) The rates of the
$nnp\Sigma$ and $pnp\Lambda$ process are of the same order of
magnitude using OPE. This result is in contrast with the case of
$W$ exchange, because the $pnp\Lambda$ process is one order of
magnitude larger than $nnp\Sigma$ process. iii) The $nnn\Lambda$
process can be included in the calculation for the bulk viscosity
using OPE.
iv) The bulk viscosity
is rather insensitive for the EoS used except with respect to the appearance of
the $\Lambda$.
\\     \indent
Therefore, the contact $W$ exchange interaction is probably to naive for quantitative
calculations.
In a more realistic OPE approach the contributions of the hyperons to the bulk viscosity are less pronounced.
\section{Acknowledgements}
The authors would like  to thank R. Timmermans and A. Ramos for
interesting discussions. This work has been supported through the
Stichting voor Fundamenteel Onderzoek der Materie with financial
support from the Nederlandse Organisatie voor Wetenschappelijk
Onderzoek.

\end{document}